\documentclass[os]{copernicus}

\usepackage{dcolumn}

\begin{document}

\begin{nolinenumbers}

\title{ENSO and the Temperature of the North Equatorial Counter Current : An Ensemble Study}

\Author[1][djw@noc.ac.uk]{David John}{Webb}
\affil[1]{National Oceanography Centre, Southampton SO14 3ZH, U.K.}

\runningtitle{ENSO and the NECC}
\runningauthor{David J. Webb}

\firstpage{1}
\maketitle

\begin{abstract}

An ensemble study using the CESM climate model is used to investigate the response of the atmosphere to changes in sea surface temperatures along the path of the North Equatorial Counter Current.  The results support those of a previous study which concentrated on a single run, confirming the sensitivity of deep convection in the central Pacific to the temperature of the counter current, and its impact on the Hadley Circulation, the Southern Oscillation and the equatorial winds in the Pacific.  A comparison with the ECMWF ERA5 data, finds that the atmospheric reanalysis fields show similar correlation with the temperature of the counter current.

\end{abstract}

\section{Introduction}

While validating a high resolution version of the Nemo ocean model, \cite{Webb_2018b} found that the North Equation Counter Current (NECC) was much warmer than normal during the development of the strong El Ni\~nos of 1982-83, 1997-98 and 2015-16.  The warm water appeared to have been advected from the West Pacific Warm Pool by a stronger than normal NECC.  The latter was due to the North Equatorial Trough (NET) being deeper than normal in the central and western Pacific during the previous months.

A study using satellite data \citep{Webb_et_al_2020} confirmed the temperature and sea level findings and \cite{Webb_2021} showed that it was the winds in the western Pacific early in the year which were responsible for the increased depth of the North Equatorial Trough in the western Pacific and its effect on NECC.  \cite{Wyrtki_1973, Wyrtki_1974}, using sea level data, reported a similar change in the depth of the trough, prior to an El Ni\~no.

The results show that the increased ocean temperatures in the central Pacific are due to processes occurring well before any other evidence that an El Ni\~no is developing.  It is thus possible that the increased temperature values are a cause of the following strong El Ni\~no, not a consequence.

More recently a study of tropical convection, using the ECMWF ERA5 data \citep{Webb_2025a}, found that deep atmospheric convection above the Pacific Inter-Tropical Convergence Zone (ITCZ) was also much stronger than normal during the development of the strong El Ni\~nos of 1982-93, 1997-98 and 2015-16.  As air from the south entering the ITCZ has to cross the warm core of the NECC, it is likely that the increased temperature of the NECC is responsible for the additional deep atmospheric convection.

This increase in deep atmospheric convection over the ITCZ is also associated with a decrease over Indonesia and south-east Asia.  As deep atmospheric convection near the Equator is responsible for the rising part of the Hadley Circulation, this change in the longitudinal structure of the Hadley Circulation has an ENSO-like impact, similar to that discussed by Hoskins \citep{Hoskins&Yang&Fonseca_2020, Hoskins&Yang_2021} when studying the movement of convection in the western equatorial Pacific.

\subsection{The NECC/ITCZ Hypothesis}

These observations led to the hypothesis that the North Equatorial Counter Current and the Inter-Tropical Convergence Zone are responsible, at least in part, for the development of the strong El Ni\~nos.  This hypothesis was tested in \cite{Webb_2025b}, using the CESM global climate model, by comparing the results from a control run, with those obtained for a forced run in which the near surface temperature of part of the NECC was increased by 1\textdegree C.  The forced run was started on the 1st August and comparisons were made between the averages of the different fields during September.

The results showed that there was a significant increase in deep atmospheric convection over the forcing region.  There was also a reduction in deep convection over the Island Continent, the region of strong diurnal convection which includes Indonesia, SE Asia and parts of Australia.  The atmospheric pressure field at sea level showed a Southern Oscillation type response with lower pressures over the western Indian Ocean and higher pressures over the eastern Pacific.  This resulted in reduced easterly winds along the Equator in the Pacific, reduced Ekman upwelling in the ocean and some increase in near surface ocean temperatures.

The comparison was repeated, for the same months, for a further ten years of a control run.  The results showed that convection in the forcing region is very sensitive to the natural fluctuations in the temperature of the NECC.  At 800~\unit{hPa}, the value of Omega, the rate of change of pressure experienced by an air particle (this is negative when convection is active), reduced from near zero at 28\textdegree C to -0.06~\unit{Pa~s^{-1}} at 30\textdegree C.  At 300~\unit{hPa}, near the top of the troposphere, the value was near zero at 29\textdegree C, and again reduced to near -0.06 near 30\textdegree C.

These results are consistent with the ERA5 results which show that during an average year significant ITCZ convection reaches the 500~\unit{hPa} level but that during strong El Ni\~nos it reaches higher in the atmosphere to 300~\unit{hPa}.

Overall the climate model results imply that temperature fluctuations of the NECC have a significant effect on atmospheric convection, the pressure field and winds, and a weaker impact on the equatorial sea surface temperatures usually associated with El Ni\~nos.  Together with the results from ERA5 and the Nemo model run, these results imply that future studies of El Ni\~no and the southern Oscillation need to seriously consider the impact of the North Equatorial Counter Current and its interaction with the Inter-Tropical Convergence Zone.

\subsection{The Present Study}

The study reported here extends the results of \cite{Webb_2025b}, using a set of 52 forced runs to study the atmospheric response in more detail, with information on both the mean response and the corresponding distribution of signal to noise. ratio.  The Southern Oscillation described by \cite{Walker_1928}, covers a similar latitude band to the Hadley Circulation, indicating that the El Ni\~no - Southern Oscillation (ENSO) is actually a mode, or a related set of modes of the Hadley Circulation itself.

For this reason the study also investigates changes in the pressure field near 200~\unit{hPa} and a north-south section through the forcing region, to learn more about the possible changes in the structure of the Hadley Circulation.  Previously studies of the effect on the Hadley Circulation, for example \cite{Hoskins&Yang&Fonseca_2020} and \cite{Hoskins&Yang_2021}, have concentrated on the effect of changes in deep atmospheric convection close to the Island Continent.  The results here show that convection in the central and eastern Pacific at 10\textdegree N, can also have a significant impact.

In the rest of the paper, section 2 briefly describes the forcing experiment and the model used for the study, and section 3 discusses use of the signal to noise ratio.  Section 4 presents the results and is followed by a final discussion section.

\section{The Model and the Forcing Experiment}

This study makes use of version 2.1.3 of CESM, the NCAR Community Earth System Model \citep{Hurrell_etal_2013, Lauritzen_etal_2018}.  The model is widely used by the Earth science community for global change research \citep{Schneider_Kay_Hannay_2022, Holland_etal_2024}, and for studies of individual parts of the climate system \citep{Li_Zhou_etal_2018, Dolores_etal_2022, Maher_etal_2023}.

The earlier study by \cite{Webb_2018b} indicated that there was a key period around September when water leaving the West Pacific Warm Pool early in the year, would be crossing the central Pacific and starting to reach the eastern Pacific.  For this reason after a long control run, a set of forced runs were carried out, starting on the 1st August each year.  Averages of the key fields were then collected for the following September and compared with the corresponding September of the control run.

In the forced runs, sea surface temperature was modified along part of the observed NECC.  The area was bound in the west and east by 180\textdegree E and 240\textdegree E and to the south and north by 4\textdegree N and 7.5\textdegree N.  Within the region surface temperature was increased by approximately 1\textdegree C over the control run.

This was done by averaging the control run temperatures on the west and east boundaries of the area on the 1st August each year.  For the forced runs, the values were increased by 1\textdegree C and the surface layers of the ocean were relaxed to a value which was a linear function of longitude between the two increased end values.

Further details about the CESM model and the forcing scheme are given in \cite{Webb_2025b}.

\section{Atmospheric Convection}

To quantify atmospheric convection, Figs. \Ref{Fig_01} and \Ref{Fig_02} show the average rate of change of atmospheric pressure following particles crossing 800~\unit{hPa} and 300~\unit{hPa} in the control runs and the average change between the forced and control runs.

The figure also shows the confidence ratio C, defined as the value of the mean change divided by the estimated error of the mean.  If there are $N$ pairs of forced and control runs, and $x_i$ is the change observed at a point in the $i$'th pair of runs, then if $M$ is the mean, $S$ the standard deviation and $C$ the confidence ratio,

\begin{eqnarray}\label{Eqn_1}
M &=& \sum_i x_i,\\
S &=& \sum_i (x_i-M)^2,\\
C &=& M/(S/(N-1))^{1/2}.
\end{eqnarray}

Values of $C$ roughly follow an $\operatorname{erfc}$ function.  If $C$ equals $1.28$, the probably that the result arises by chance is $10\%$.  Values of $1.64$, $2.05$ and $2.32$ correspond to probabilities of $5$\%, $2$\% and $1$\%.  These values are usually described as corresponding to significance of 90\%, 95\%, 98\% and 99\%.

Figures in this paper showing the signal to noise ratios, also show the sign of the ratio instead of the absolute value normally displayed.  This is done to emphasize the sign of the most significant changes.

\begin{figure*}[t]
  \begin{center}
    \includegraphics[width=14cm,viewport=35 100 550 700, clip]{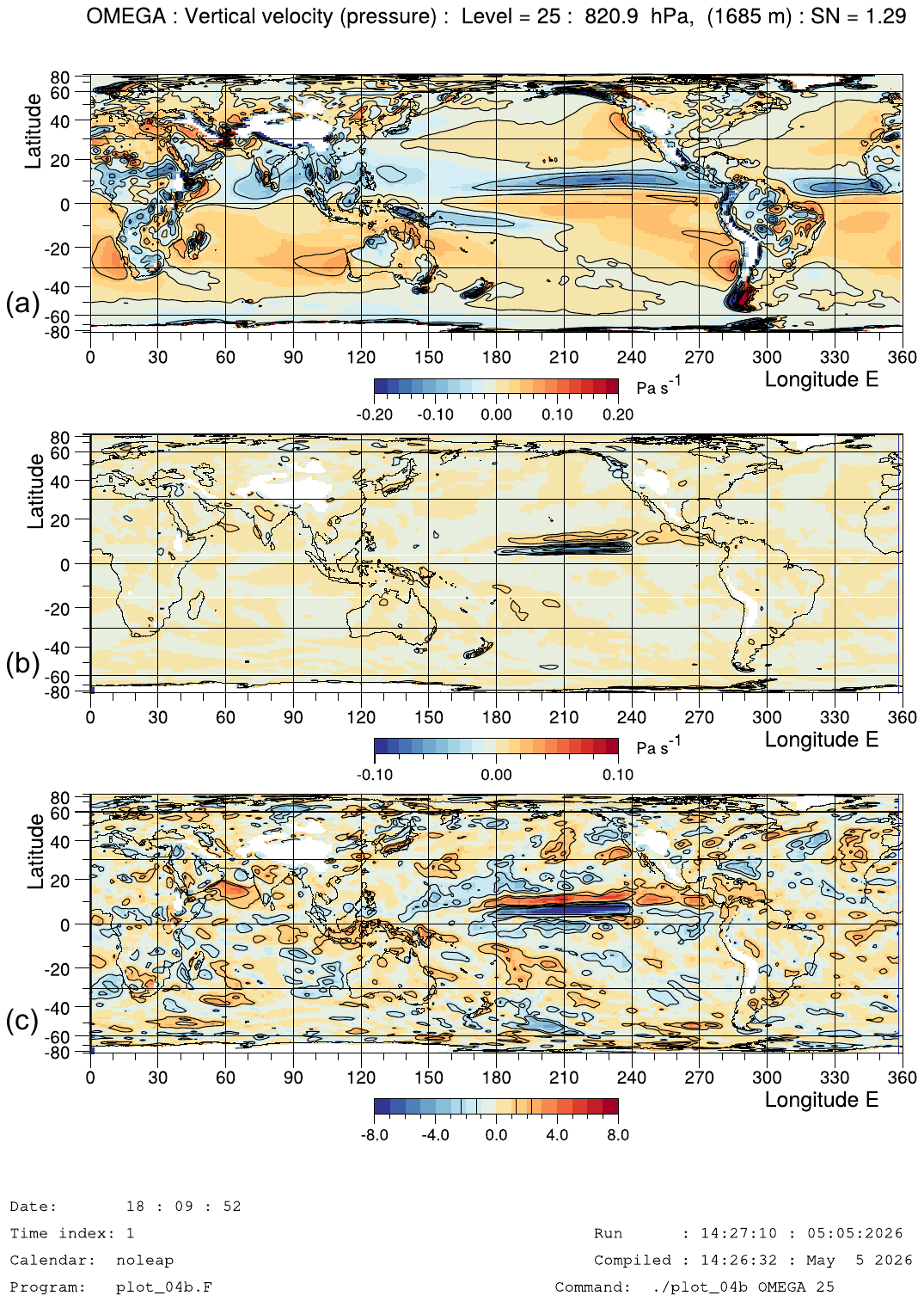}
  \end{center}
  \caption{\label{Fig_01} (a) Value of Omega at 1685 m (around 820 hPa) averaged over each September of the control run.  Contours at intervals of 0.05~\unit{Pa~s^{-1}}.  Negative (blue) values correspond to convection. (b) Difference between the forced and control runs averaged over the same months.  The forcing, described in the text, is non-zero within the marked rectangular region in the central Pacific.  Contours at intervals of 0.01~\unit{Pa~s^{-1}}.  (c) Signal to noise ratio defined in Eqn. \ref{Eqn_1}.  The contours correspond to 90\% and 99\% probability that the values are significant and not due to random background noise. }
\end{figure*}

\begin{figure*}[t]
  \begin{center}
    \includegraphics[width=14cm,viewport=40 100 550 700, clip]{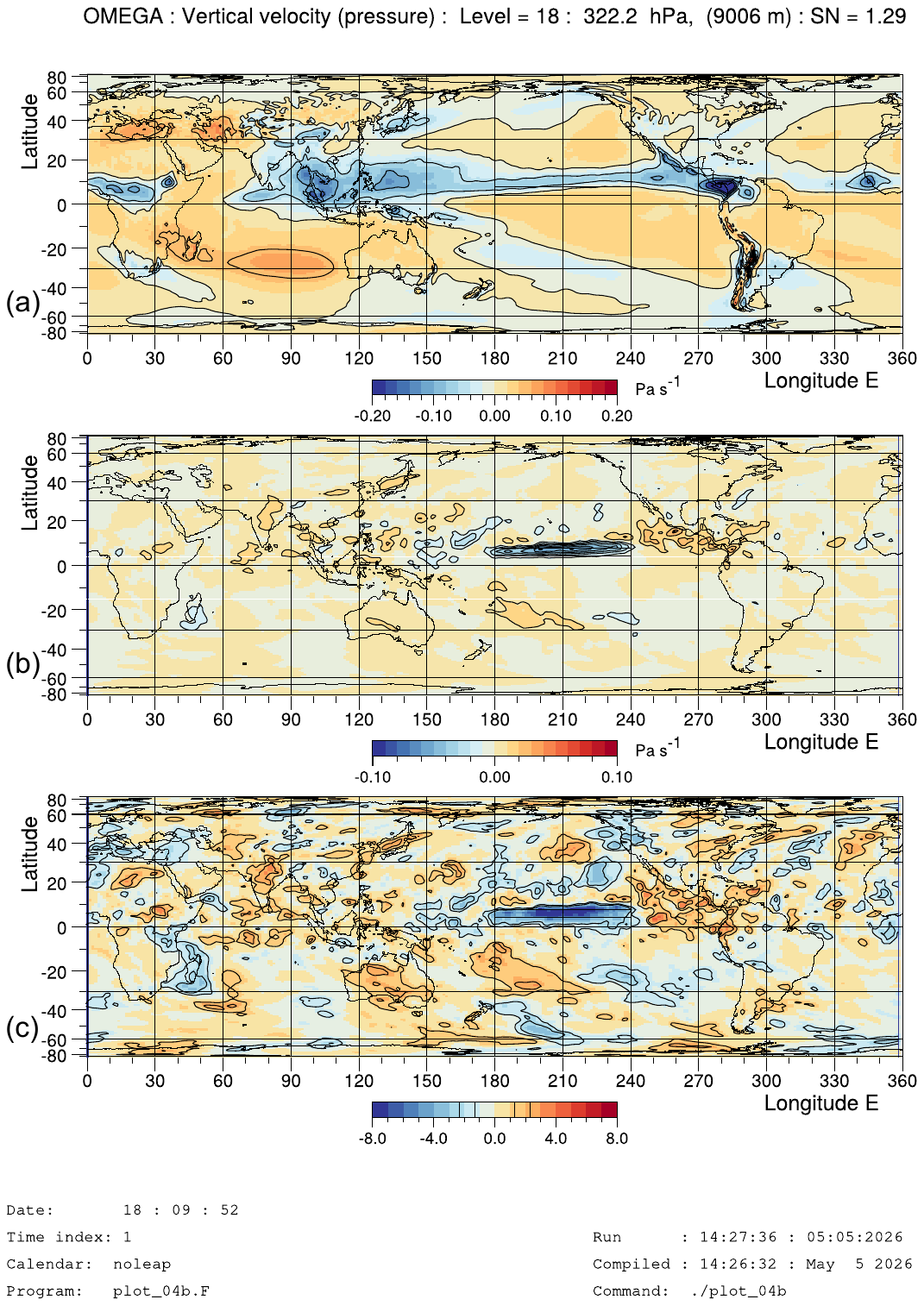}
  \end{center}
  \caption{\label{Fig_02} (a) Value of Omega at 9000 m (around 322 hPa) averaged over each September of the control run.  Contours at intervals of 0.05~\unit{Pa~s^{-1}}.  Negative (blue) values correspond to convection. (b) Difference between the forced and control runs averaged over the same months.  The forcing, described in the text, is non-zero within the marked rectangular region in the central Pacific.  Contours at intervals of 0.01~\unit{Pa~s^{-1}}.  (c) Signal to noise ratio defined in Eqn. \ref{Eqn_1}. The contours correspond to 90\% and 99\% probability that the values are significant and not due to random background noise.  }
\end{figure*}

\begin{figure*}[t]
  \begin{center}
    \includegraphics[width=14cm,viewport=10 275 550 790, clip]{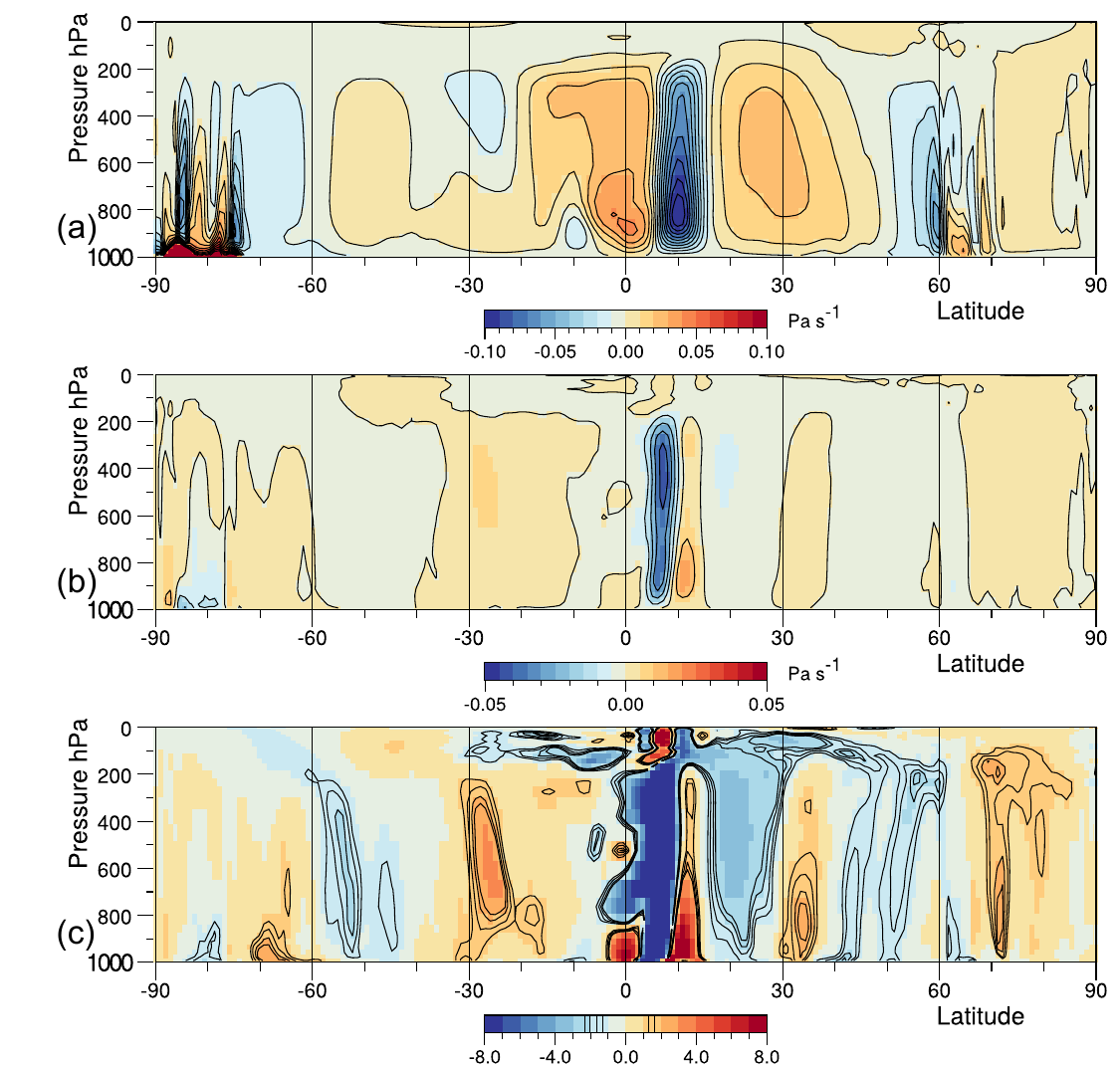}
  \end{center}
  \caption{\label{Fig_03} (a) Value of Omega in the control run averaged over the longitudes between 180\textdegree E and 240\textdegree E (120\textdegree W) and averaged over each September of the control run.  Contours at intervals of 0.01~\unit{Pa~s^{-1}}.  Negative (blue) values correspond to convection. (b) Difference between the forced and control runs averaged over the same months.  Contours at intervals of 0.01~\unit{Pa~s^{-1}}.  (c) Signal to noise ratio defined in Eqn. \ref{Eqn_1}.The contours correspond to 90\%, 95\%, 98\% and 99\% probability that the values are significant and not due to random background noise.  }

\end{figure*}

\section{Convection at 800 and 300~\unit{hPa}}

At 800~\unit{hPa} (Fig. \Ref{Fig_01}a) the strongest convection occurs in the central and eastern equatorial Pacific, near 10\textdegree N, and in the corresponding region of the Atlantic.  When SST forcing is applied in the central Pacific (Fig. \Ref{Fig_01}b) the main difference is an increase of convection in the forcing region and a reduction in convection further north.  This has similarities with the southward movement of convection in the western equatorial Pacific observed during El Ni\~nos \citep{Rasmusson&Carpenter_1982}.

The signal to noise ratio, Fig. \Ref{Fig_01}c, shows that in the area of SST forcing. the signal to noise ratio is high, indicating that the southward movement of the convection feature has a very high level of significance.  Elsewhere the areas with an apparent large signal to noise ratio tend to be small in area and randomly scattered, as one would expect from noise.  However when such areas with similar sign are closely grouped, as occurs with the areas of reduced convection between the forcing region and Central America, their net contribution is probably significant.

At 300~\unit{hPa} (Fig. \Ref{Fig_02}a), the control run shows that convection over the Pacific and Atlantic ITCZ is much weaker than at 800~\unit{hPa}.  Instead there is increased convection over the Island Continent north of the Equator and off central America.  There are also changes over the Indian Ocean and Africa.

In the forced runs the forcing region again shows increased convection, but at 300~\unit{hPa} there is no associated region of reduced convection to the north.  As discussed in \cite{Webb_2025b}, normally a large fraction of ITCZ convection only reaches the middle atmosphere, around 500~\unit{hPa}, but here the small increase in temperature in the forcing region is enough to allow most of the convection to reach the 300~\unit{hPa} level.  Such a change is similar to that found by \cite{Webb_2025a} in the ERA5 data when comparing strong El Ni\~no years with normal years.

The signal to noise ratio is very large in the forcing region, but elsewhere the patchy nature again indicates that most of the apparent correlations are noise.  However again, when such areas are closely grouped, the changes may be significant.

\section{Meridional Section}

Figure \Ref{Fig_03}a shows the value of Omega, averaged over the forcing longitudes (180\textdegree E to 240\textdegree  E) and over each September of the control run.  It shows that, in the central Pacific, there is normally significant convection up to 600 hPa.  Above this convection drops off with height, with a small amount reaching 200~\unit{hPa}.  There is sinking over the Equator, again with a maximum below 600~\unit{hPa} and a further region of sinking air near 30\textdegree N.  The latter is the normally accepted limit of the Hadley Circulation.

When SST forcing is added, Fig. \Ref{Fig_03}b, there is increased convection, above the warm water near 8\textdegree N, at all heights up to 200~\unit{hPa}.  Matching this, there is a region of reduced convection near 11\textdegree N, the result of a southward movement of the main latitudes of convection.

Within the convective column the main increase occurs above 600~\unit{hPa}, the result of the increased depth of convection discussed in the last section.  At other latitudes the changes are marginal, except near 30\textdegree S, where there is a slight reduction in the upward movement of air in the middle atmosphere.  Atmospheric physics in this region is complex, involving the interaction between the tropical circulation of the dynamically active jet stream.

Although many of the changes are small, Fig \Ref{Fig_03}c shows that after averaging over longitudes, as has been done here, many of them are statistically significant.  Near the Equator the changes to ITCZ convection have very high significance, but there is also high significance associated with the reduced sinking between 15\textdegree N and 30\textdegree N, and the increased sinking at 30\textdegree S.

These high correlations are further evidence that the SST forcing is affecting the large scale Hadley Circulation.  Other areas of high correlation nearer the poles indicate that deep atmospheric convection in the tropics has significant impacts on high latitudes in both hemispheres.

\section{Fluctuations in SST}

\cite{Webb_2025b} found that atmospheric convection above the NECC was very sensitive to small changes in SST.  The result was based on a relative small sample, so to confirm that the result is still valid with a much larger set of runs, Fig. \Ref{Fig_04} shows the average values of Omega for each September of the control and forced runs, as a function of SST at 800~\unit{hPa} and 300~\unit{hPa}.

As before the figure shows the effect of noise due to the random events occurring within each run, but in addition there is a significant increase in deep atmospheric convection with temperature.  The linear fit to the distribution is, at 800~\unit{hPa},
\begin{eqnarray}\label{Eqn_8}
\Omega_{800} &=& -0.0257*(T_{SST}-28.27) ,
\end{eqnarray}
where $T_{SST}$ is the average sea surface temperature.  At 300~\unit{hPa},
\begin{eqnarray}\label{Eqn_9}
\Omega_{300} &=& -0.0258*(T_{SST}-28.68) .
\end{eqnarray}
The slope of the two lines are essentially the same, the main difference being the $0.4$\textdegree C difference before the convection reaches 300~\unit{hPa}.

At 300~\unit{hPa} the linear relationship also appears to break down above 30\textdegree C, with the average convection rate being greater than that expected for the linear relation.  At 800~\unit{hPa} there may be a similar effect, but this is masked by a greater variability in the convection data.

At temperatures below 28.5\textdegree C at 300~\unit{hPa} and 28.3\textdegree C at 800~\unit{hPa}, the results are dominated by the contribution of sinking air.  Maximum values lie in the range $0.02$ to $0.03$~\unit{Pa~s^{-1}}, except for the two cases where the SST values are lowest.  In these two runs the values at 300~\unit{hPa} lie in the same range, but lower in the atmosphere, at 800~\unit{hPa}, the sinking rates are much larger.

\subsection{Comparison with the Cold Pool}

In order to provide a better understanding of the low temperature fluxes, Fig. \ref{Fig_05} shows the corresponding results for  the Cold Pool region,  where the contribution from convection is expected to be a minimum.  The two figures show that for SST values between $19$\textdegree C and $27$\textdegree C the value of Omega at both levels lies between $0.015$ and $0.030$~\unit{Pa~s^{-1}}.

Physically this is consistent with temperatures changes being dominated by cooling due to upward radiation into space.  Although there will be some warming by radiation coming from the ocean surface, the effect of the temperatures changes (299\textdegree K to 310\textdegree K) will have little effect compared with the imbalance of radiation with space (3\textdegree K).  For the air at $300$~\unit{hPa}, a sinking rate of $0.025$~\unit{Pa~s^{-1}}, implies a period of around 30 days before the air reaches the ocean surface.

\begin{figure*}[t]
  \begin{center}
    \includegraphics[width=14cm,viewport=30 175 760 490, clip]{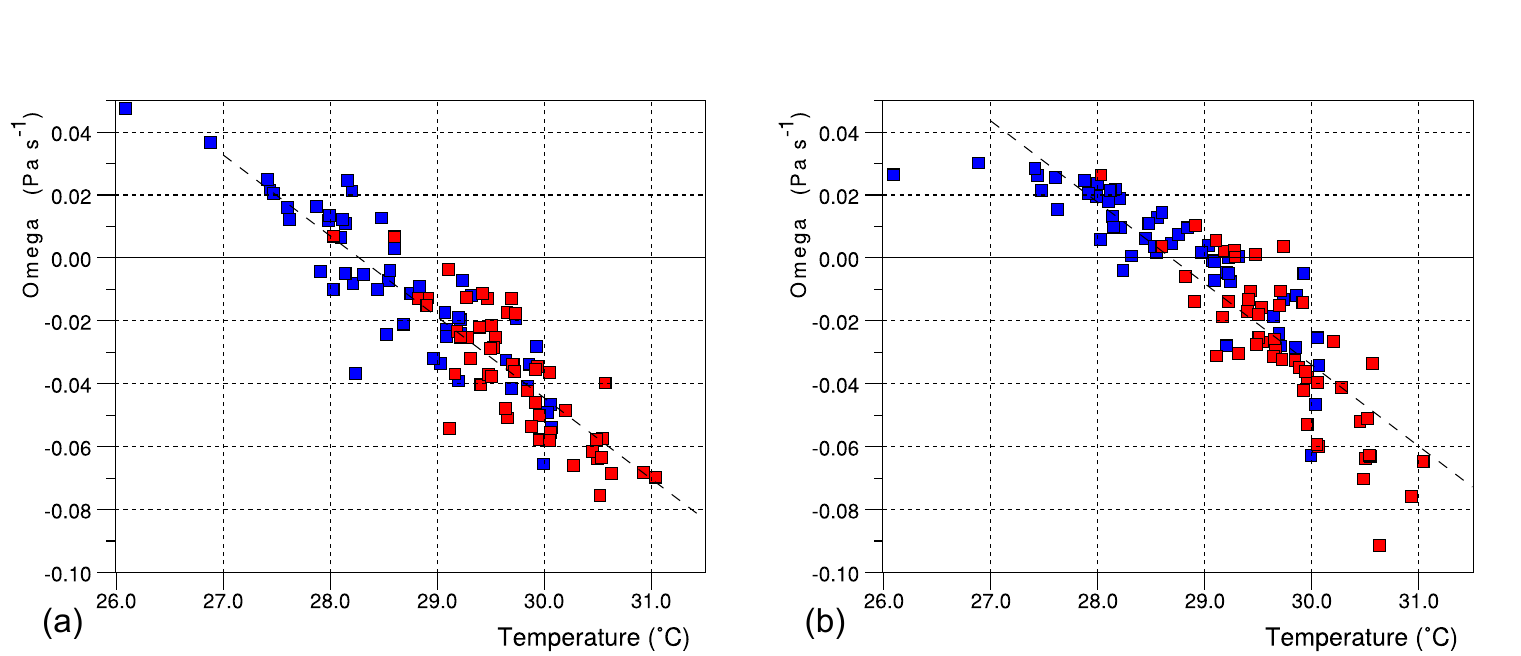}
  \end{center}
  \caption{\label{Fig_04} Average values of Omega above the forcing region during each September of the control (blue) and forced (red) runs at (a) 800~\unit{hPa} and (b) 300~\unit{hPa}.}
\end{figure*}

\begin{figure*}[t]
  \begin{center}
    \includegraphics[width=14cm,viewport=30 175 760 490, clip]{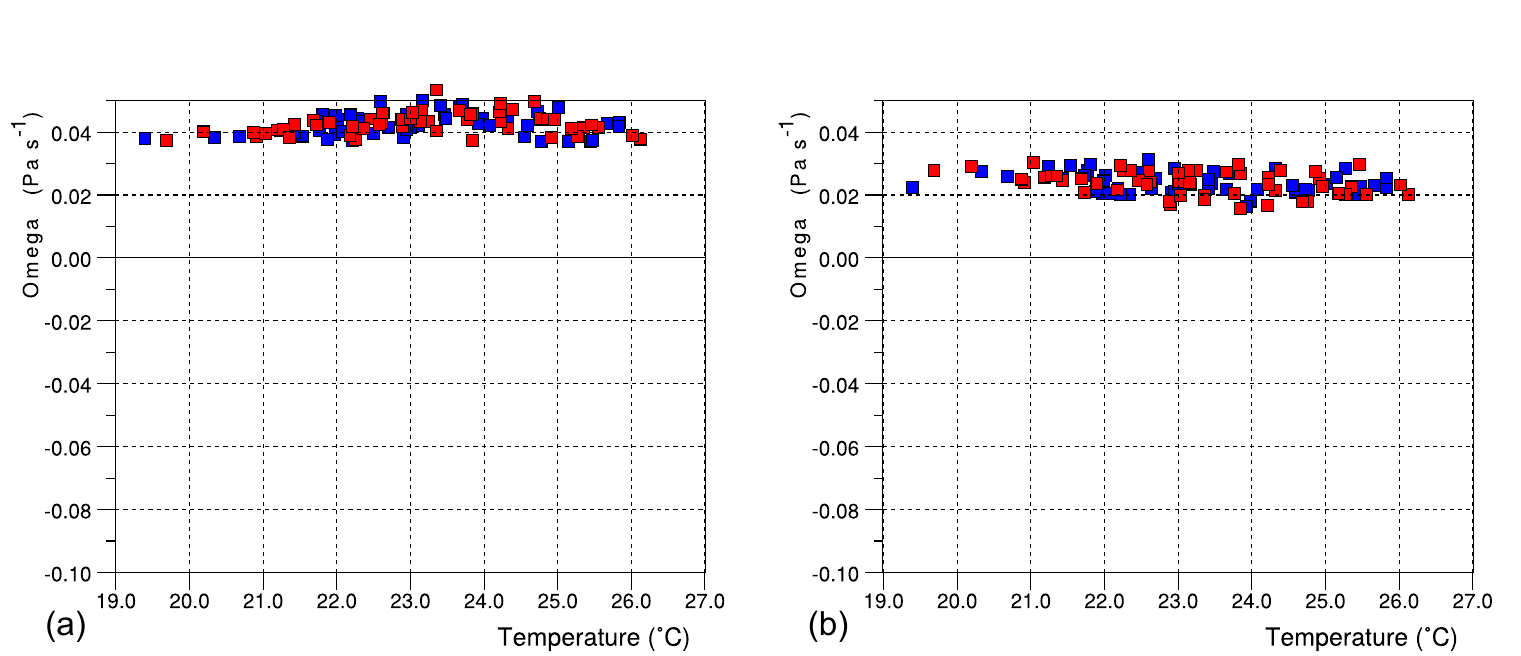}
  \end{center}
  \caption{\label{Fig_05} Average values of Omega above the Cold Pool (130\textdegree E-260\textdegree E, 3\textdegree S-3\textdegree N) region during each September of the control (blue) and forced (red) runs at (a) 800~\unit{hPa} and (b) 300~\unit{hPa}.}
\end{figure*}

\subsection{Tropical Convection}

The results summarized by Fig. \ref{Fig_04} and \ref{Fig_05} show that Omega is dominated by radiational sinking below $27$\textdegree C, but that above this value there is a rapid increase of convection with temperature.

The behaviour is consistent with descriptions of tropical atmospheric convection illustrated and discussed by \cite{Riehl_1979} (Fig 2.23), \cite{Gill_1982} (Fig. 3.5) and \cite{Bechtold_2019} (Fig. 1.24).  These are based on the results presented by \cite{Jordon_1958} on the average vertical structure of the tropical atmosphere during GATE, the GARP Atlantic Tropical Experiment.

Jordon's data shows that the equivalent potential temperature of the surface air, i.e. its potential temperature plus the heat content due to the humidity, needs to exceed $350$\textdegree K before tropical convection can start.  Below this limit the lower layers of the atmosphere prevent convection developing.  Above the limit, convection can start but only when some large scale convergence raises the near surface air to the condensation level.

The air then has the potential to rise to the top of the troposphere but usually mixing with the relative dry air of the middle atmosphere limits the height reached.  It is only after a series of such events have increased the humidity of the middle atmosphere that deep convection can eventually occur.

The representation of such a complex sub-grid scale process in an atmospheric model is a challenging process, and so it is possible that some of the results presented here have systematic errors.  As stated above convection depends on the equivalent potential temperature, not the actual surface temperature.  However in the temperature range being considered, saturated water content is increase by about 7\% for $1$\textdegree C increase in temperature, and as near surface humidity stays around $70$\% to $80$\%, changes in equivalent potential temperature will be primarily due to the surface temperature changes.

As a result the (almost) hard temperature limit on convection, shown by the model, is realistic.  Once such a limit has been passed, convection depends on random events and it is likely that the success rate of such events will increase with temperature as seen here, resulting in an approximately linear dependency.  However the slope of the line and any further non-linearity may depend on the details of the atmospheric model being used.

\begin{figure*}[t]
  \begin{center}
    \includegraphics[width=14cm,viewport=100 110 700 570, clip]{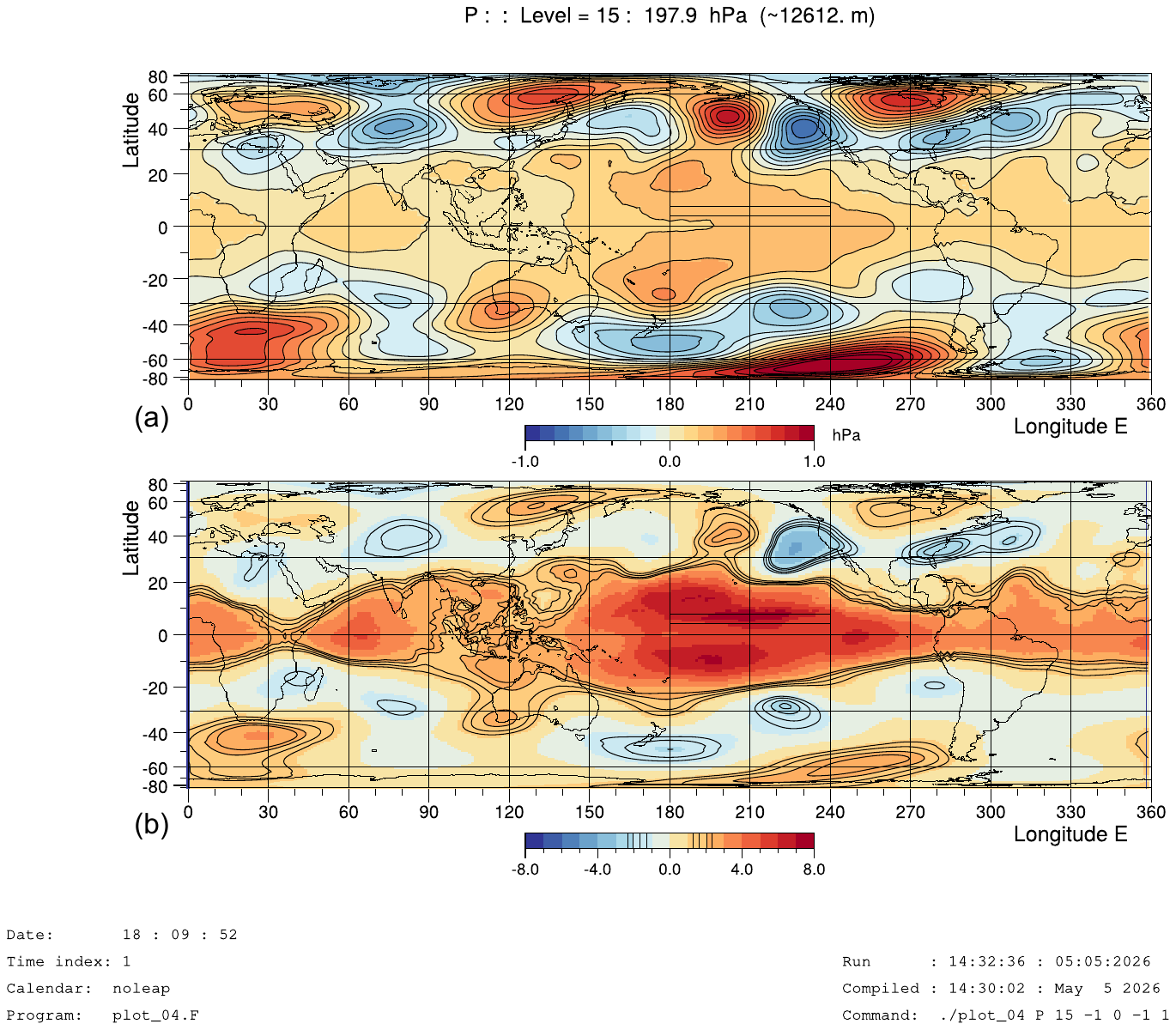}
  \end{center}
  \caption{\label{Fig_06} (a) Difference in atmospheric pressure at 12612~\unit{m} ($\sim$200~\unit{hPa}) between the forced and control runs, when averaged over all Septembers. Contours at intervals of $0.1$~\unit{hPa}.  (b) Signal to noise ratio defined in Eqn. \ref{Eqn_1}. The contours correspond to 90\%, 95\%, 98\% and 99\% probability that the result is significant and not due to random background noise. }
\end{figure*}

\begin{figure*}[t]
  \begin{center}
    \includegraphics[width=14cm,viewport=100 110 700 570, clip]{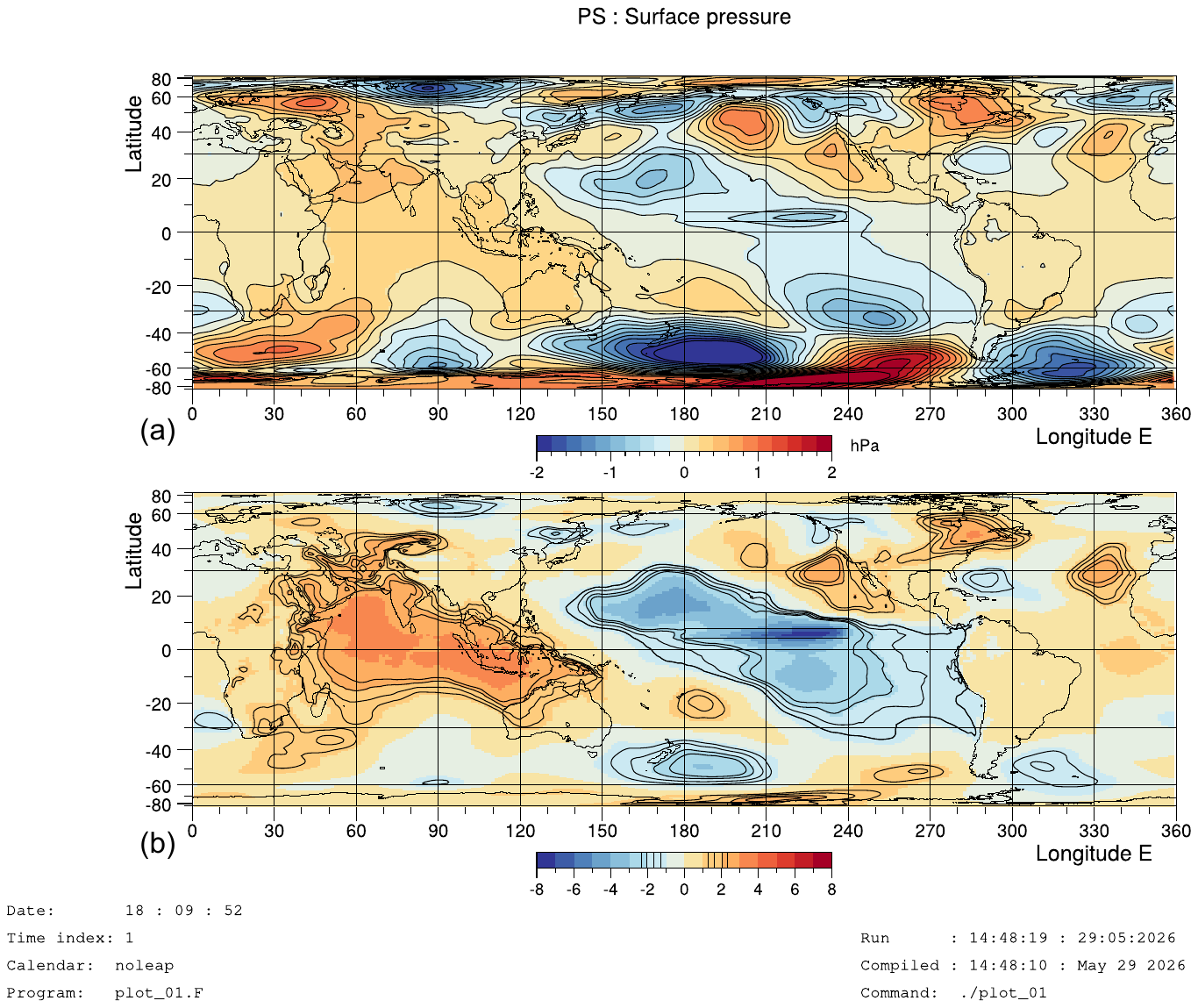}
  \end{center}
  \caption{\label{Fig_07} (a) Difference in sea surface atmospheric pressure between the forced and control runs, when averaged over all Septembers.  Contours at intervals of $0.1$~\unit{hPa}. (b) Signal to noise ratio defined in Eqn. \ref{Eqn_1}. The contours correspond to 90\%, 95\%, 98\% and 99\% probability that the result is significant and not due to random background noise. }
\end{figure*}

\section{Pressure Changes near the Tropopause}

Figure \ref{Fig_06}, shows the pressure change around 200~\unit{hPa} due the the forcing.  Changes are only of order $0.2$~\unit{hPa} but in the tropical band there are large areas where the signal to noise ratio is greater than three.

Over the Pacific main tropical disturbance extends to $30$\textdegree north and south, confirming as with Fig. \ref{Fig_03}, that convection over the SST forcing region is modifying the Hadley Circulation.  Although the forcing lies north of the Equator,  the perturbation pressure field is, to first order, symmetric about the Equator.

Along the Equator, the maximum pressure change lies just to the west of the forcing region, but two areas where the change is greater lie north and south of the Equator at $180$\textdegree E.  Other areas with both increased pressure and high significance occur over the Indian and Atlantic Oceans.  Other more localized regions with significant amplitude and significance occur at high latitudes.  Features that may be of particular interest include the positive and negative anomalies in the North Pacific, the negative anomaly over the Gulf Stream region and the two positive anomalies in the Southern Ocean, one near Drake Passage and one south of Africa.

The pattern of pressure changes shown over the Pacific between $30$\textdegree S and $30$\textdegree N, has interesting similarities to the patterns reported by \cite{Gill_1980} in his analytic single vertical mode studies of the response of the tropical atmosphere to surface heating at the Equator.  When the source was symmetric about the Equator, his solution consisted of an eastward traveling Kelvin wave on the  Equator plus two Rossby waves to the west of the forcing region, one north and one south of the Equator.  All three waves are first mode baroclinic with low pressures near the surface and high pressures near the tropopause, as found here.

Gill also considered a case of forcing by a source north of the Equator.  He found that the Rossby wave to the north is enhanced and that to the south is weaker and of opposite sign.  This differs from the pattern seen in Fig \ref{Fig_06}, where the southern Rossby wave is the stronger of the two.

Despite this difference, the similarities between the two numerical and analytic results indicate that the response of the Hadley Cell to the sea surface temperature perturbation may be approximately linear and can be explained in terms of the generation of Kelvin and Rossby waves by the convecting plumes.

\section{Sea Level Pressure and the Southern Oscillation}

Figure \ref{Fig_07} shows the average change in sea level pressure due to the forcing and the signal to noise ratio.  As discussed in \cite{Webb_2025b} the pressure pattern has strong similarities with that expected for the Southern Oscillation \citep{Rasmusson&Carpenter_1982}.  Pressures are low in the eastern Pacific, high in the Indian Ocean, with the zero line extending from the western Pacific, near 30\textdegree N, to the central Pacific near 30\textdegree S.

As with the Southern Oscillation pressure field, there is an anomalous pattern of high and low pressure in the north-east Pacific, near 230\textdegree E, 40\textdegree N, that may be associated with the path the jet stream takes in crossing the Rockies.  In the south the jet stream passes south of South America, so the interaction with the Hadley cell may be responsible for the large changes (but low probability) generated close to Drake Passage.

Other features that may be of note include the low pressure region associated with New Zealand and high and low pressure regions of the North Atlantic.  These may be due to a set of Rossby waves resulting from the changes in the Hadley circulation over the Pacific.

\begin{figure*}[t]
  \begin{center}
    \includegraphics[width=14cm,viewport=100 110 700 570, clip]{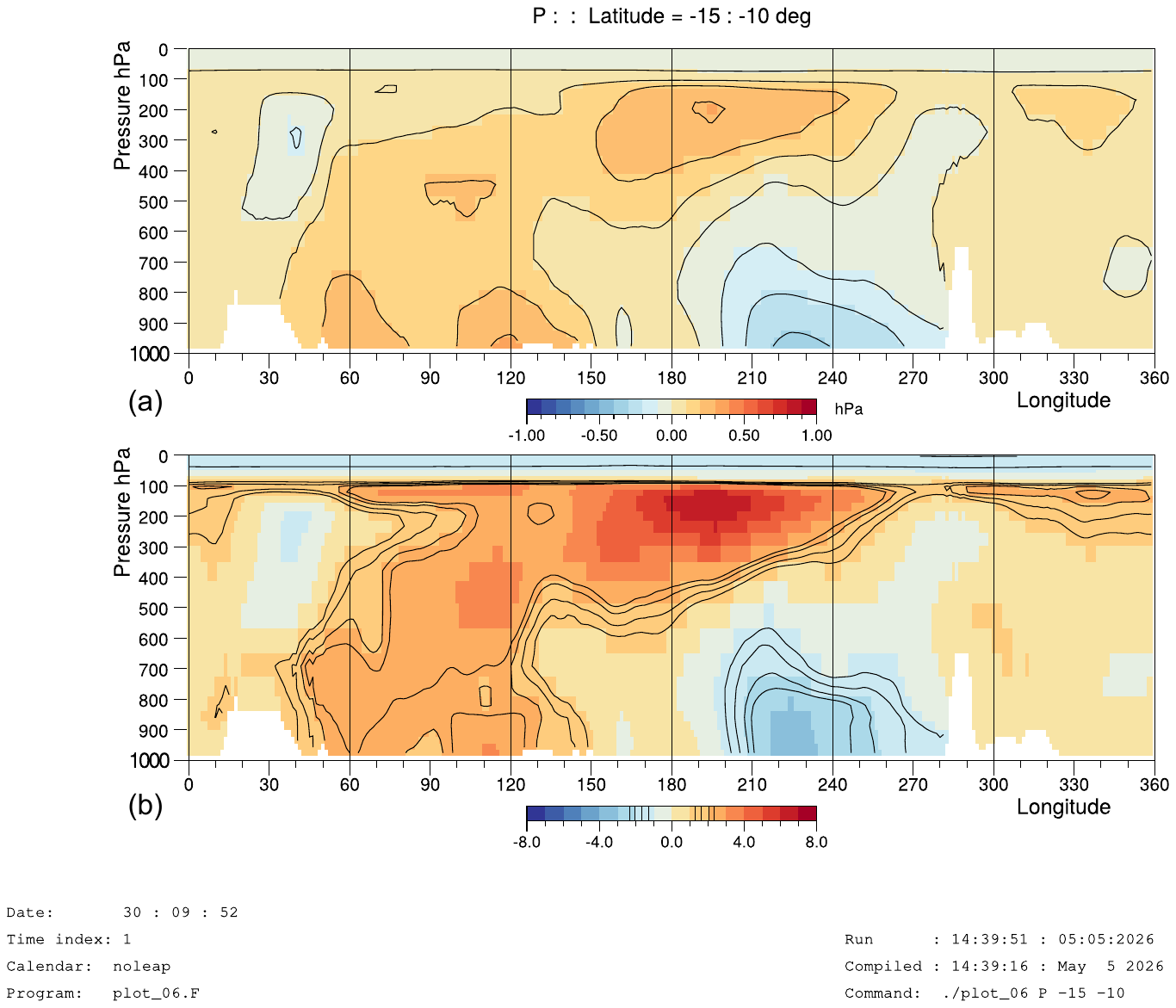}
  \end{center}
  \caption{\label{Fig_08} (a) (a) Difference in atmospheric pressure between the forced and control runs, when averaged over all Septembers and over latitudes between 15\textdegree S and 10\textdegree S.  Contours at intervals of $0.1$~\unit{hPa}.  (b) Signal to noise ratio defined in Eqn. \ref{Eqn_1}. The contours correspond to 90\%, 95\%, 98\% and 99\% probability that the values are significant and not due to random background noise. }
\end{figure*}

\begin{figure*}[t]
  \begin{center}
    \includegraphics[width=14cm,viewport=100 110 700 570, clip]{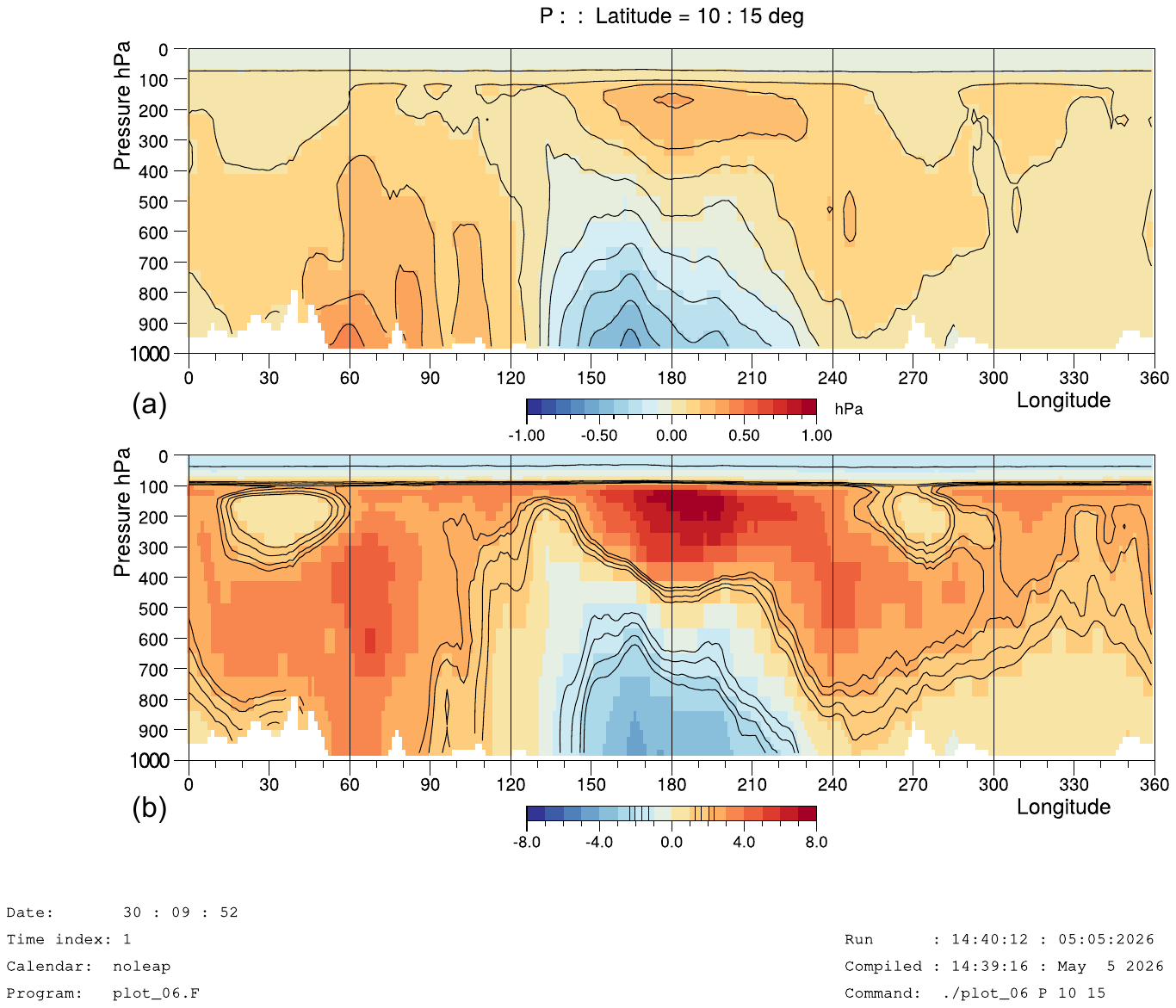}
  \end{center}
  \caption{\label{Fig_09} (a) Difference in atmospheric pressure between the forced and control runs, when averaged over all Septembers and over latitudes between 10\textdegree N and 15\textdegree N.  Contours at intervals of $0.1$~\unit{hPa}.  (b) Signal to noise ratio defined in Eqn. \ref{Eqn_1}. The contours correspond to 90\%, 95\%, 98\% and 99\% probability that the values are significant and not due to random background noise. }
\end{figure*}

\begin{figure*}[t]
  \begin{center}
    \includegraphics[width=14cm,viewport=100 97 700 570, clip]{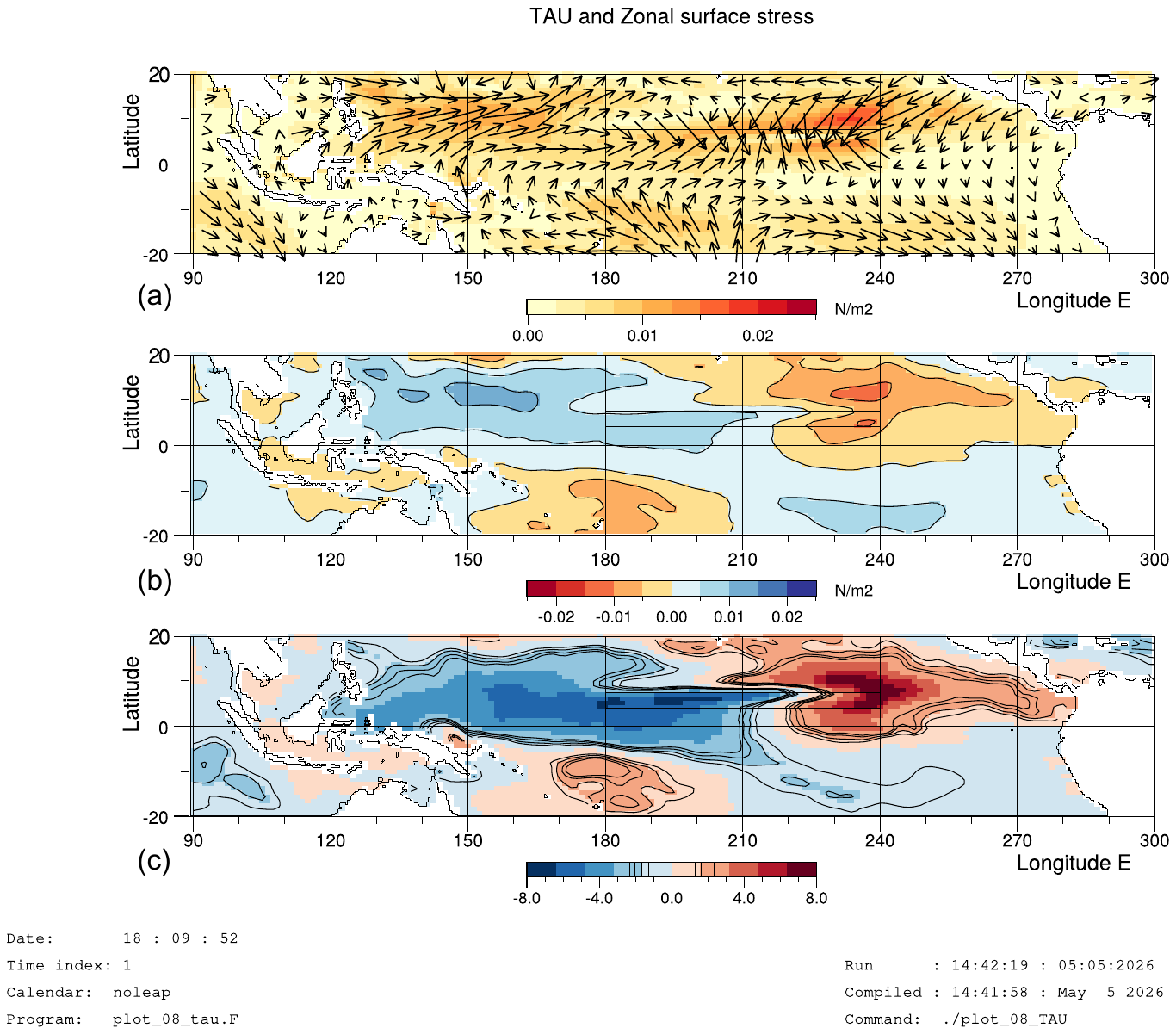}
  \end{center}
  \caption{\label{Fig_10} (a) Vector difference between the forced and control runs in the wind stress acting on the ocean, when averaged over all Septembers.  (b) Average change, due to the forcing, in the zonal component of the stress.  Contours at intervals of $0.005$~\unit{N m^{-2}} (c) Signal to noise ratio defined in Eqn. \ref{Eqn_1}. The contours correspond to 90\%, 95\%, 98\% and 99\% probability that the values are significant and not due to random background noise. }
\end{figure*}

\section{Zonal Sections North and South of the Equator}

The results presented above have shown that the SST perturbation of the NECC has generated a large scale response both near the sea surface and high in the atmosphere.  The response high in the atmosphere can be explained in part by the analysis of \cite{Gill_1980}.  However the Southern Oscillation pattern seen at sea level is so different from the pattern high in the atmosphere, that some additional physical process must be involved.

An important clue to the additional process can be obtained by considering the zonal distribution of pressure changes.  In Fig. \ref{Fig_08}, this is done as a average between 15\textdegree S and 10\textdegree S.  Above $400$~\unit{hPa} the figure shows the high pressures associated with the Rossby wave discussed earlier.  However whereas as in a single mode analytic model this would be mirrored by a low pressure region near the surface, in the more realistic numerical model the low pressure region is shifted towards the east Pacific and the Andes.

One of the difficulties in trying to understand the Hadley Circulation is to explain how the overturning circulation flows away from the Equator high in the atmosphere and towards the Equator close to the surface.  If the flow is geostrophic then at levels where integrals can be continued uninterrupted around the Earth, the northward and southward flows must balance, so the net flow must be zero.

High in the atmosphere the solution appears to be that the flow has a non-geostrophic component  \citep{Vallis_2017}.  This allows the air to move away from the Equator, conserving its angular momentum about the Earth's axis by increasing its eastward velocity until it eventually interacts with the jet streams.

At low levels Hadley's original suggestion that it was equatorial flow was made possible by the force of the trade winds on the ocean.  However Fig. \ref{Fig_08} shows that another mechanism is also active, involving the pressure differences that develop across the Andes and the highlands of Africa.  These pressure differences allow net north-south transports at all heights within the lower atmosphere.

In the Indian and Pacific Ocean section, the responses indicates that, at heights up to $850$~\unit{hPa} there is a net northward flow between 150\textdegree E and 200\textdegree E, with geostrophic balance recirculation to the west, extending to Africa and east to the Andes.  Above this level there can be a continuing northward flow balanced by the pressure difference across the Andes.

The corresponding plot north of the Equator is shown in Fig. \ref{Fig_09}.  The topography is much lower, but again there is net flow towards the Equator at heights up to 800~\unit{hPa}, mainly between $90$\textdegree E and $110$\textdegree E.

\subsection{Near Surface Pressures and ENSO}

These observations imply that the Southern Oscillation is a result of variations in the strength of the return branch of the Hadley Circulation, driven by fluctuations in the total amount of deep atmospheric convection nearer the Equator.  In the forced experiments this takes the form of an inflow due to the SST induced additional convection.

In the normal pattern of ENSO events \citep{Trenberth&Caron_2000}, the sign of the relationship between the El Ninos and the Southern Oscillation indicates that El Ni\~nos are associated with flows towards the Equator and implied enhanced convection, as seen here.

The changes in pressure gradient across the Pacific and Indian Ocean close to the Equator will also affect pressure at the Equator.  In the east, the Andes prevent zonal winds at low level and, as a result, the north-south pressure gradient along the coast must be close to zero.  In the west, boundary winds, such as the East African Jet will be driven by a north-south pressure gradient, but as seen in Fig. \ref{Fig_07}, their overall effect is small.

As a result, the unexpected large scale changes in the pressure gradient along the Equator, seen in the forced runs, appear to be primarily the result of the inflows caused by increased deep convection near the Equator.  As the pressure gradient affects winds and Cold Pool upwelling, the mechanism probably explains much of the observed connection between the Southern Oscillation and El Ni\~no \citep{Trenberth&Caron_2000}.

\section{Winds near the Equator}

Figure \ref{Fig_10} shows the difference in the stress acting on the ocean surface between and forced and control runs, .  Around the region of the SST forcing, the vectors of Fig \ref{Fig_10}a, show a convergence towards the convection region.  They also show some evidence of air circulation around the area of convection.  This process is also seen in the plot of the zonal component of stress, which shows stronger easterlies to the north of the forcing region and weaker easterlies to the south.  The figure also shows the effect of the more northerly winds, south of the Equator between 180\textdegree E and 210\textdegree E, reflecting the gradient in surface pressure seen in Fig \ref{Fig_08}.

On the Equator, the zonal component of stress shows that there is a region of enhanced easterlies near 240\textdegree E, associated with the inflow to the forcing region, and a larger region of reduced easterlies to the south and west.  In a number of the comparison runs this this region of reduced easterlies appears to have extended to areas east of New Guinea.

Together with the evidence of the relationship between the equatorial pressure gradient and the Southern Oscillation, the results imply that the winds and wind stresses acting along the Equator are affected by two mechanisms.  On the large scale, convection within the equatorial band increases the low level geostrophic balanced inflow, generates a Southern Oscillation response, which reduces the average pressure gradient and average strength of the easterlies along the Equator.  In addition, each convective event can generate a local response, increasing the easterlies to the east and reducing the easterlies or increasing the westerlies to the west.

\begin{figure*}[t]
  \begin{center}
    \includegraphics[width=14cm,viewport=100 97 700 570, clip]{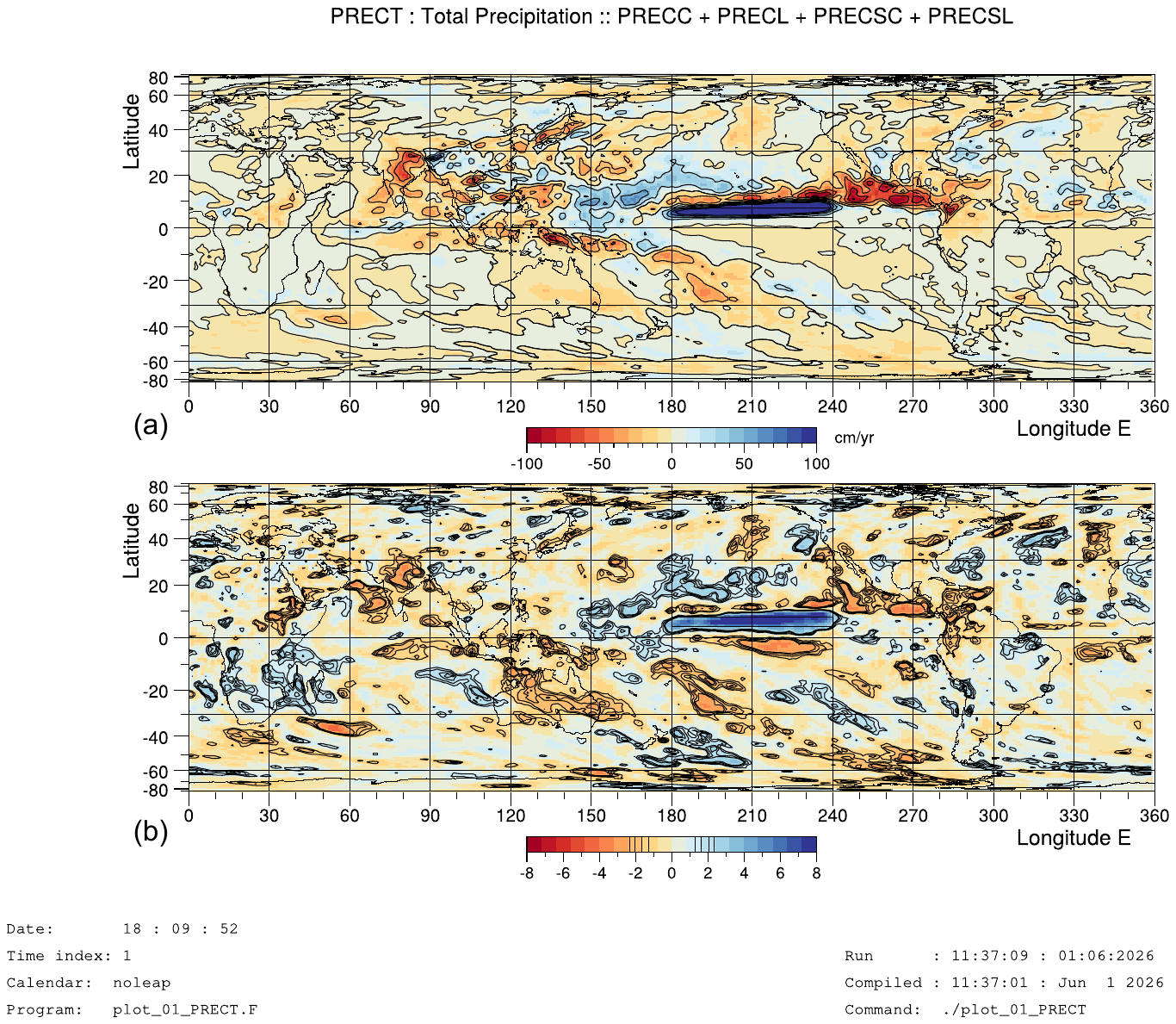}
  \end{center}
  \caption{\label{Fig_10a} (a) Difference in rainfall between the forced and control runs averaged over all Septembers.  (b) Signal to noise ratio defined in Eqn. \ref{Eqn_1}. The contours correspond to 90\%, 95\%, 98\% and 99\% probability that the values are significant and not due to random background noise. }
\end{figure*}

\section{Teleconnections}

The earlier discussion of surface atmospheric pressures concentrated on the Southern Oscillation signal which dominates the Pacific and Indian Oceans in the equatorial band.  However Fig. \ref{Fig_07} also shows high probability features which may reflect steering of the jet streams.  In the north the pairs of features at 230\textdegree E, 290\textdegree E and 340\textdegree E form a regular pattern which may include the high pressure region near 45\textdegree E.

The pattern is likely to be a result of meanders in the jet stream.  Individually the regions of high pressure have the largest signal to noise ratio and, except in the western Atlantic, the low pressure regions have low probability.  However the overall pattern is very unlikely to be the result of background noise.  In the eastern Atlantic the pattern has strong similarities to that expected from the North Atlantic Oscillation.

The Southern Hemisphere shows no corresponding extended pattern, however the region of low pressure south-east of New Zealand has a very high signal to noise ratio.  The high and low pressure regions that lie to the west and east of Drake Passage both have significance probabilities of over 95\%.

\subsection{Rainfall}

Figure \ref{Fig_10a} shows the average change in rainfall between the control and forced runs.  The larges changes in rainfall are found over the forcing regions and this is also the region with the highest signal to noise ratio.  Elsewhere the rainfall plot and the signal-to-noise plot are both dominated by small scale features.

Unfortunately many of these features may be the result of random noise due to weather, but where there is a more extensive pattern of such features, there is a greater chance that the general behaviour can be considered more seriously.  An example of this is the reduction in rainfall that occurs just north of the forcing region, indicating a southward shift of the ITCZ.  A second example is the much larger reduction in rainfall that occurs further east, in a region that extends to Central America and into the Caribbean.

Other features include the band of increased rainfall extending from the western equatorial Pacific towards North America and an irregular region of lower rainfall extending from India, through Indonesia and New Guinea into the South Pacific Convergence Zone (SPCZ). \cite{VanderWeil_etal_2016} found that a reduction in the SST temperature gradients across the Pacific would also reduce rainfall in the SPCZ.

Returning to the present study, India, whose rainfall stimulated Walker's original study \citep{Walker_1928} shows a reduction of over 60~\unit{cm y^{-1}} with probabilities of over 99\%.  The low rainfall regions of central Australia are also affected with a high probability of a further small reduction in rainfall.

\section{Discussion}

The study made use of 52 different forced runs, each starting at the beginning of August in one of the years of a control run.  In each of the forced runs the sea surface temperature along the path of the North Equatorial Counter Current in the central Pacific was increased by $1$\textdegree C relative to the temperature of the control run.

Key variables were then averaged over the following month, and the differences from the control run used to determine the mean effect of the forcing and the signal to noise ratio.  The final results support those of \cite{Webb_2025b} and provide additional insights into the physics behind the response of the coupled system.

\subsection{Changes in Convection}

The study shows that increased SST along the line of the NECC, moves the band of ITCZ convection closer to the Equator.  Observations show a similar movement of the ITCZ during both east Pacific \citep{Rasmusson&Carpenter_1982} and central Pacific El Ni\~nos \citep{Xie&Yang_2014}.  The present model results imply that, in both types of El Ni\~no, the increased temperature is a cause, probably the major cause, of the southward movement of the ITCZ.

The present results show that the surface temperature forcing increases ITCZ convection at the $800$~\unit{hPa} and $300$~\unit{hPa} levels, and that these are separated by regions of detrainment and entrainment in the middle atmosphere.  The results also show some reduction in deep atmospheric convection over the Island Continent and South-east Asia.

At the longitudes of the forcing, there are changes in the vertical mass fluxes near $30$\textdegree N and $30$\textdegree S.  Together these results indicate that the forced convection is affecting both the longitude and latitude structure of the Hadley Circulation.

The results confirm the finding of \cite{Webb_2025b}, that deep atmospheric convection is sensitive to the regular fluctuation in temperature of the NECC.  The results of Appendix A, also show that vertical fluxes in the atmosphere are insensitive to the much larger range of SST values associated with the Pacific Cold Pool.

Near the $200$~\unit{hPa} level, the increased convection, in the forced runs, results in a pressure anomaly pattern covering most of the Pacific between $30$\textdegree N and $30$\textdegree S.  The signal to noise ratio for the feature is high.  The increased convection also affects pressures over the Indian and Atlantic Oceans, with a lower signal to noise ratio, and results in some large changes in pressure at high latitudes.

The pattern of the perturbation over the Pacific is of interest because it closely follows the pattern of a Kelvin wave plus two Rossby waves, found by \cite{Gill_1980} in his single baroclinic mode study of equatorial convection.  Although in the present case the forcing is away from the Equator, the atmospheric response is, to first order, similar to that of equatorial forcing.

\subsection{The Southern Oscillation and Geostrophy}

In Gill's analytic model, the surface response is the opposite to that found high in the atmosphere.  However the present results show that in the real world this is not the case, the large scale response at sea level instead being similar to that of the Southern Oscillation, with a sign corresponding to that of an El Ni\~no.  Thus it has low pressures over the central, south-west and north-west Pacific, with high pressures over the Indian Ocean and near the North American coastline.   Low pressures in the central Pacific are also enhanced by an additional region of low pressures around the forcing region.

The reason for the decoupling between the high and low levels of the atmosphere can be understood from zonal sections made just outside the equatorial band.  These show the presence of high topography which intersects the lower atmosphere.  Pressure differences can develop across the topography and these allow geostrophic winds to generate net transports towards and away from the Equator.  Such flows are not possible in the upper atmosphere because of the lack of meridional boundaries.  In the present forced case, the negative Southern Oscillation index allows an inflow of air towards the Equator to replace that lost by the forced convection.

The decoupling between the lower and upper atmosphere, implies that the pattern of pressure changes found at $200$~\unit{hPa} is due to the presence of a high vertical mode of the atmosphere or a series of such modes.  This helps to explain why after a month, the atmospheric Kelvin wave is still close to the forcing region.  As was found in the Southern Ocean by \cite{Webb&deCuevas_2007} and \cite{DoosWebb_1994}, the fast low vertical mode Kelvin waves can propagate away, leaving north-south flows balanced by pressure differences across topography.  The high order vertical modes remain, leaving the pressure anomalies seen high in the atmosphere.

\subsection{Sea level Pressure and Winds}

The presence of topography also constrains the sea level pressure gradient along the eastern and western boundaries of the ocean close to the Equator.  As a result, the equatorial convection and the resulting negative Southern Oscillation index, results in a reduced large scale pressure gradient along the Equator.  This reduces the strength of the easterlies along the Equator.

An additional change in the winds caused by the local response to the convection forced by the changes in SST.  The results show that this increases the strength of the easterlies in the east equatorial Pacific.  It reduces their strength in the west and so, together with the large scale reduction in the strength of the easterlies, may sometimes result in a region of westerlies.

\subsection{NECC-ITCZ and ENSO : An Hypothesis}

The present study has confirmed the strong coupling between the NECC and ITCZ, and confirmed how small changes in the temperature of the NECC can generate large ENSO like changes to the Hadley Circulation.  Previously \cite{Webb_2018b} showed that warm water from the western Pacific Warm Pool was responsible for the extreme SST temperatures observed prior to strong El Ni\~nos.  The study also showed that it takes six to nine months for the warm water to cross the Pacific.

\cite{Meyers_1979} showed that at 6\textdegree N, Rossby waves take a similar time to propagate across the Pacific.  The waves are generated by fluctuations in ITCZ convection and were found by \cite{Webb_2018b} to affect the speed of the NECC in the central and western Pacific.  Such changes will affect the temperatures along the path of teh NECC.

The time scales associated with advection and Rossby wave propagation are both comparable with those seen in ENSO observations.  They mean that there is a large disconnect between temperature changes in the Western Pacific Warm Pool and their impact, via the NECC and ITCZ, on the Hadley circulation and ENSO.

Warm water carried by the NECC may be cooled by tropical instability waves, but with this exception, the large time delays in the transport of heat across the Pacific mean that it is reasonable to hypothesize that the NECC-ITCZ contribution to ENSO develops independently of any other process contributing to ENSO.

\begin{figure*}[t]
  \begin{center}
    \includegraphics[width=14cm,viewport=30 175 760 460, clip]{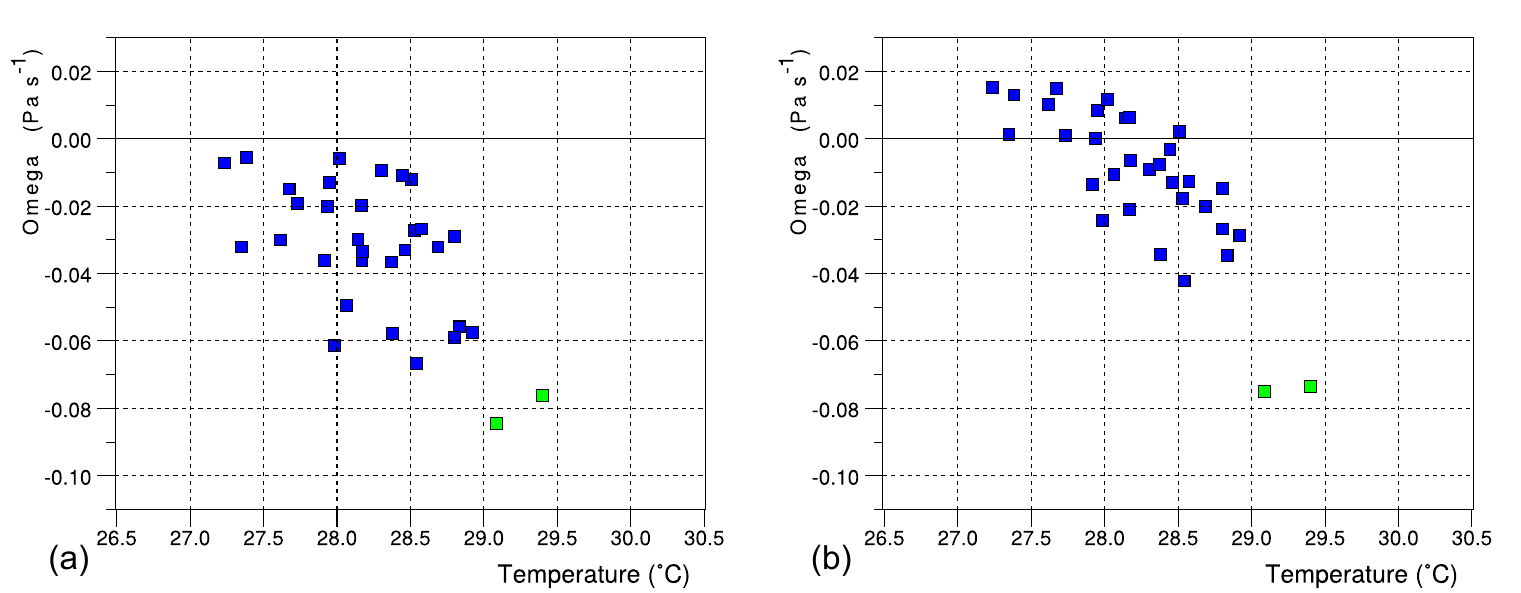}
  \end{center}
  \caption{\label{Fig_11} ERA5 estimates of Omega, at heights of (a) 800~\unit{hPa} and (b) 300~\unit{hPa}, above the forcing region during each September during the years 1990 to 2200.  The green symbols correspond to 1997 and 2017 during the growth of strong El Ni\~nos.}
\end{figure*}

\begin{figure*}[t]
  \begin{center}
    \includegraphics[width=14cm,viewport=30 175 760 460, clip]{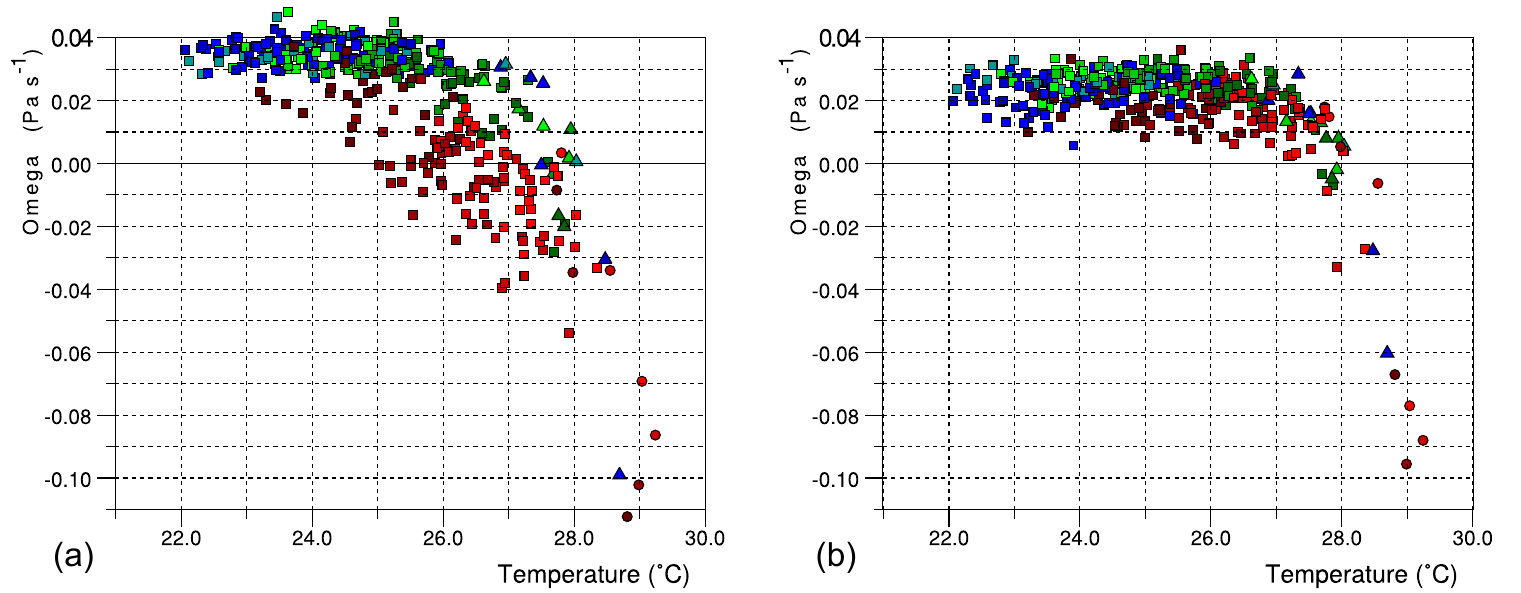}
  \end{center}
  \caption{\label{Fig_12} ERA5 estimates of Omega, at heights of (a) 800~\unit{hPa} and (b) 300~\unit{hPa}, above the cold pool (230\textdegree E-270\textdegree E, 3\textdegree S-3\textdegree N) for each month during the years 1990 to 2020.  Colours grade from blue for each January to red for each December.  Months just prior to a strong El Ni\~no are marked with circles, the months just after by triangles.}
\end{figure*}

\appendix

\section{ERA5 Data}

As a check on the model results, Figs. \ref{Fig_11} and \ref{Fig_12} use ERA5 data from the period 1990 to 2020 to plot the value of Omega over the forcing and cold pool regions, at 800~\unit{hPa} and 300~\unit{hPa}.

Figure \ref{Fig_11},which covers the forcing region, corresponds to Fig. \ref{Fig_04} of the present paper, plotting the monthly average values only for the September of each year.  It shows that at both 800 and 300~\unit{hPa}, Omega has a roughly linear dependence on temperature.  In both cases the slope of the curve is similar but the zero crossing point is about one degree cooler in the ERA5 data than in the model.

Figure \ref{Fig_12} is for the Cold Pool, and includes results every month between 1990 and 2020, with colour being used to represent the position of each month during the year.  At 800~\unit{hPa}, the figure shows that Omega is usually independent of temperature, as was found for the model in September.  However in November and December, temperatures tend to be warmer than average.  The lower atmosphere is also more unstable so that in some cases Omega becomes negative, indicating net convection.

At 300~\unit{hPa}, the seasonal cycle in SST can be seen but this has little effect on Omega.  Exceptions only occur around the times of the two strong El Ni\~nos in 1997/98 and 2015/16m when temperatures are above 27\textdegree C .

Overall the observations support those of the model, and indicate that deep atmospheric convection over the NECC is sensitive to  its fluctuations in temperature and that the flux of sinking air over the Cold Pool is, to first order, independent of temperature.

\section{ENSO Theory}

The results of this paper diverge in a number of respects from the results expected on the basis of some widely accepted theories of ENSO.

1.  \citeauthor{Bjerknes_1966} (\citeyear{Bjerknes_1966}, \citeyear{Bjerknes_MWR_1969}) description of the possible response of the Hadley Circulation to equatorial sea surface temperature, proposed an overturning equatorial ``Walker Circulation''  with convection in the western equatorial Pacific and sinking air over the ``Cold Pool'' of the central and eastern equatorial Pacific.  His Walker Circulation  was limited to the north and south by the cloud bands of the two Inter-Tropical Convergence Zones.

Bjerknes proposed that this circulation was separate from the Hadley Circulations of the Northern and Southern Hemispheres but that, near the tropopause, changes in the strength of the Walker Circulation  somehow modified the momentum transfer between the Equator and the mid-latitude jet streams.

The present study does not show this to be wrong but it does show that convection in the ITCZ also needs to be considered when discussing changes in the Hadley Circulation due to ENSO.

2.  \cite{Bjerknes_MWR_1969} proposed that the strength of the Walker Circulation was proportional to the sea surface temperature difference between the east and west equatorial Pacific.  In practice SSTs in the western Pacific only vary by a degree or so, whereas in the Cold Pool they can fluctuate by over six degrees.  Thus, to first order, Bjerknes's model circulation is driven by the large fluctuations of temperature in the east.  Other models either use the same approximation as Bjerknes or are based explicitly on the temperature of the Cold Pool \citep{Battisti&Hurst_1989, Jin_1997}.

Figures \ref{Fig_05} and \ref{Fig_12} show that, in both the model and the ERA5 reanalysis, deep atmospheric convection over the Cold Pool is essentially independent of temperature.  Thus the rate of sinking in this branch of the Walker circulation does not  directly respond to surface temperature.  The sinking rate can be modified by occasional convective events, the ERA5 data showing that at low levels in the atmosphere this does happen near the end of each year.  However this is not enough to explain the pattern of El Ni\~nos and La Ni\~nas and their global impact.

3.  The size of the Cold Pool does change with Cold Pool temperature.  As a result, if Bjerknes's temperature proposition is replaced by one involving Cold Pool area, then during a La Ni\~na, when the Cold Pool area is high, the total amount of sinking over the Cold Pool and the resulting strength of the Walker Circulation and its convective branch, should be high.

However the present results connecting the Southern Oscillation index with the low level inflow towards the Equator, indicate that the total amount of convection is highest during El Ni\~nos, and lowest during La Ni\~nas.

4.  The month to month changes in the strength of winds along the Equator are normally attributed to the changing strength of the Walker Circulation.

The present results imply that the large scale changes in the pressure gradient along the Equator is caused by the Southern Oscillation, itself a consequence of changes in the Hadley Circulation and changes in the strength of deep atmospheric convection near the Equator.

The large scale changes will affect the mean winds along the Equator.  However, locally, the position of active convective sites can also have an effect.  As an example convection along the path of the NECC in the central Pacific, increases the strength of equatorial easterlies further east and reduces, possibly reversing, their impact in the west.

\begin{acknowledgements}
The present study was carried out at the National Oceanography Center, UK, and funded by the Natural Environment Research Council grant NE/Y005589/1.  I would like to acknowledge the generous support of NOC and its staff, especially the members of the Global Climate group.

I would also like to acknowledge previous support, especially that concerned with my interest in the equatorial Pacific, by the UK Institute of Oceanographic Sciences and the CSIRO Division of Fisheries and Oceanography.

The study made use of version 2.1.3 of the CESM2 climate model.  The model is made available and supported by the U.S. National Center for Atmospheric Research under the sponsorship of the National Science Foundation. Thanks to all the scientists, software engineers and administrators who contributed to the development of CESM2.

The ERA5 atmospheric reanalysis data was generated by the European Centre for Medium-Range Weather Forecasts and provided for analysis by the Copernicus Climate Change and Atmosphere Monitoring Services \citep{Hersbach_etal_2022}.

\end{acknowledgements}

\codeavailability{The CESM model is available from NCAR.  A copy of the modified ocean model subroutine used for the forced run is available from ``https://github.com/djwebb/ccode``.}

\bibliographystyle{copernicus}
\bibliography{Pacific_short.bib}

\end{nolinenumbers}

\end{document}